\documentclass[a4paper]{article}
\begin{document}

\title{Macroconstraints from Microsymmetries}%
\author{Ludwik
Turko\\ {\em Institute of Theoretical Physics, University of
Wroclaw,}\\ {\em pl. Maksa Borna 9, 50-204 Wroc\l aw, Poland}\\
{\small E-mail: turko@ift.uni.wroc.pl}}
\date{Talk presented at July 27, 2000}
 \twocolumn[
 \maketitle
\begin{abstract}
 The dynamics governing the evolution of a many body system
is constrained by a nonabelian local symmetry. We obtain explicit
forms of the global macroscopic condition assuring that at the
microscopic level  the evolution respects the overall symmetry
constraint.\vspace{1cm}
\end{abstract}
]

Let us consider a multiparticle quantum system with the local
interaction invariant with respect to the internal symmetry group
$G$. We shall call this underlying symmetry of microscopic
interactions a\\ \underline{microsymmetry} of the system. The
system transforms under a given representation of the symmetry
group. We shall call this property a \underline{macrosymmetry} of
the system. Let us try to answer a question: is macrosymmetry
preserved during a time evolution of the system? This problem is
treated with more details by J.~Rafelski and the
author\cite{TurRaf}.

The system consists of particles belonging to multiplets of the
symmetry group. One denotes
$f^{(\alpha_i,\nu_i)}_{(\zeta)}(\Gamma,\vec r,t)$ a distribution
function of the particle, transforming under $\alpha_i$
representation of the symmetry group, with quantum numbers of
$\nu_i$ member of the multiplet. The variables $(\Gamma,\vec r)$
denotes a  set of the phase - space variables such as $(\vec
p,\vec r)$ and $t$ is time. A subscript $\zeta$ denotes other
quantum numbers characterizing different multiplets of the same
representation $\alpha$.

The number of particles of the specie $\{\alpha,\nu_\alpha\}$ is:
\begin{equation}
N^{(\alpha)}_{\nu_\alpha;(\zeta)}(t)=\int\,dV d\Gamma
f^{(\alpha,\nu_\alpha)}_{(\zeta)}(\Gamma,\vec r,t).
\label{numb}
\end{equation}

Let us consider the corresponding state vector in particle number
representation: $\left\vert N^{(\alpha_1)}_{\nu_{\alpha_1}},\dots,
N^{(\alpha_n)}_{\nu_{\alpha_n}}\right\rangle .$ All other
variables, related to phase-space properties of the system are
suppressed here. This vector describes symmetry properties of our
systems and transforms as a direct product representation of the
symmetry group $G$. This representation is of the form:
\begin{equation}\label{prod}
\alpha_1^{N^{(\alpha_1)}} \otimes \alpha_2^{N^{(\alpha_2)}}
\otimes \cdots \otimes \alpha_n^{N^{(\alpha_n)}}.
\end{equation}
A multiplicity $N^{(\alpha_j)}$ of the representation $\alpha_j$
in this product is equal to a number of particles which transform
under this representation:
\begin{equation}
N^{(\alpha_j)} = \sum_j\left(\sum_{\zeta_j}\,
N^{(\alpha_j)}_{\nu_{\alpha_j};(\zeta_j)}\right)= \sum_j\,
N^{(\alpha_j)}_{\nu_{\alpha_j}} .
 \label{pnumber}
\end{equation}
The representation given by Eq. (\ref{prod}) can be decomposed
into direct sum of irreducible representations $\Lambda_k$.
Corresponding states are denoted as $\left\vert \Lambda_k,
\lambda_{\Lambda_k}; {\mathcal N}\right\rangle$ where
$\lambda_{\Lambda_k}$ is an index numbering members of the
representation $\Lambda$ and ${\mathcal N}$ is a total number of
particles
\begin{equation}
{\mathcal N}=\sum_k\,
N^{(\alpha_k)}_{\nu_{\alpha_k}}.
\label{number}
\end{equation}
Each physical state can be decomposed into irreducible
representation base states $ \left\vert \Lambda_k,
\lambda_{\Lambda_k}; {{\mathcal N};\xi_{\Lambda_k}}\right\rangle
$. Variables $\xi_{\Lambda}$ are degeneracy parameters required
for the full description of a state in the "symmetry space". Let
us consider a projection operator ${\mathcal P}^{\Lambda}$ on the
subspace spanned by all states transforming under representation
$\Lambda$.

\begin{eqnarray}
{\mathcal P}^{\Lambda}\left\vert
N^{(\alpha_1)}_{\nu_{\alpha_1}},\dots,
N^{(\alpha_n)}_{\nu_{\alpha_n}}\right\rangle\nonumber\\[4pt] =
\sum_{\xi_{\Lambda}}\!^{\oplus} \left\vert \Lambda,
\lambda_{\Lambda};\xi_\Lambda\right\rangle{\mathcal
C}^{\Lambda,\lambda_{\Lambda}}_{\{N^{(\alpha_1)}_{\nu_{\alpha_1}},
\dots,\,N^{(\alpha_n)}_{\nu_{\alpha_n}}\}}(\xi_\Lambda).
\label{proj1states}
\end{eqnarray}

This operator has the generic form (see e.g. \cite{Wigner}):
\begin{equation}\label{proj1}
{\mathcal
P}^{\Lambda}=d(\Lambda)\int\limits_G\,d\mu(g)\bar\chi^{(\Lambda)}(g)U(g).
\end{equation}
Here $d(\Lambda)$ is the dimension of the representation
$\Lambda$, $\chi^{(\Lambda)}$ is the character of the
representation $\Lambda$, $d\mu(g)$ is the invariant Haar measure
on the group, and $U(g)$ is an operator transforming a state under
consideration. We will use the matrix representation:
\begin{eqnarray}
&&U(g)\left\vert N^{(\alpha_1)}_{\nu_{\alpha_1}},\dots,
N^{(\alpha_n)}_{\nu_{\alpha_n}}\right\rangle\nonumber\\[4pt] &=&
\sum\limits_{\nu_1^{(1)},\dots,\nu_n^{(N_{\nu_n})}}\,
D^{(\alpha_1)}_{\nu_1^{(1)}\nu_1}\!\!\cdots
D^{(\alpha_1)}_{\nu_1^{(N_{\nu_1})}\nu_1}\!\!\cdots
D^{(\alpha_n)}_{\nu_n^{(1)}\nu_n}\nonumber\\[4pt]
&&\cdots D^{(\alpha_n)}_{\nu_n^{(N_{\nu_n})}\nu_n} \left\vert
N^{(\alpha_1)}_{\nu_{\alpha_1}},\dots,
N^{(\alpha_n)}_{\nu_{\alpha_n}}\right\rangle .
 \label{transf}
\end{eqnarray}
$D^{(\alpha_n)}_{\nu,\nu}$ is a matrix elements of the group
element $g$ corresponding to the representation $\alpha$. Notation
convention in Eq.\,(\ref{transf}) arises since  there are
$N^{(\alpha_j)}_{\nu_{\alpha_j}}$ states transforming under
representation $\alpha_j$ and having quantum numbers of the
$\nu_{\alpha_j}$-th member of a given multiplet.

The probability
\[\overline{P^{\Lambda,\lambda_{\Lambda}}_{\{N^{(\alpha_1)}_{\nu_{\alpha_1}},
\dots,\,N^{(\alpha_n)}_{\nu_{\alpha_n}}\}}}\] that
$N^{(\alpha_1)}_{\nu_{\alpha_1}},\dots, N^{(\alpha_n)}_{
\nu_{\alpha_n}}$ particles transforming under the symmetry group
representations $\alpha_1,\dots,\alpha_n$ combine into $\mathcal
N$ particle state transforming under representation $\Lambda$ of
the symmetry group is given by
\begin{eqnarray}
&\left\langle N^{(\alpha_1)}_{\nu_{\alpha_1}},
\cdots,N^{(\alpha_n)}_{\nu_{\alpha_n}}\right\vert {\mathcal P
}^{\Lambda}\left\vert N^{(\alpha_1)}_{\nu_{\alpha_1}},\dots,
N^{(\alpha_n)}_{\nu_{\alpha_n}}\right\rangle\nonumber\\[4pt] &=
\sum\limits_{\xi_\Lambda}\vert{\mathcal
C}^{\Lambda,\lambda_{\Lambda}}_{\{N^{(\alpha_1)}_{\nu_{\alpha_1}},
\dots,\,N^{(\alpha_n)}_{\nu_{\alpha_n}}\}}(\xi_\Lambda)\vert^2 .
\label{normP}
\end{eqnarray}

Left hand side of this equation can be calculated directly from
Eqs.(\ref{proj1}) and (\ref{transf}). One gets finally
\begin{eqnarray}
 &&\overline{P^{\Lambda,\lambda_{\Lambda}}_{\{N^{(\alpha_1)}_{\nu_{\alpha_1}},
\dots,\,N^{(\alpha_n)}_{\nu_{\alpha_n}}\}}}\nonumber\\[4pt]
&=&{\mathcal A}^{\{{\mathcal N}\}}
d(\Lambda)\int\limits_G\,d\mu(g)\bar\chi^{(\Lambda)}(g)
[D^{(\alpha_1)}_{\nu_1\nu_1}]^{N^{(\alpha_1)}_{\nu_{\alpha_1}}}\nonumber\\[4pt]
&&\cdots
[D^{(\alpha_n)}_{\nu_n\nu_n}]^{N^{(\alpha_n)}_{\nu_{\alpha_n}}} .
\label{weights}
\end{eqnarray}
where ${\mathcal A}^{\{{\mathcal N}\}}$ is a permutation
normalization factor. For particles of the kind $\{\alpha,\zeta\}$
we included in Eq.\,(\ref{weights}) the permutation factor:
\begin{equation}\label{permfac1}
{\mathcal A}^\alpha_{(\zeta)}= \frac{{\mathcal
N}^{(\alpha)}_{(\zeta)}!}{\prod\limits_{\nu_\alpha} {\mathcal
N}^{(\alpha)}_{\nu_\alpha;(\zeta)}!} .
\end{equation}
The permutation factor ${\mathcal A}^{\{{\mathcal N}\}}$ is a
product of all "partial" factors
\begin{equation}\label{permfactor}
{\mathcal A}^{\{{\mathcal N}\}}=
\prod\limits_j\prod_{\zeta_j}{\mathcal A}^{\alpha_j}_{(\zeta_j)}.
\end{equation}

Because of macrosymmetry all weights in Eq.\,(\ref{weights}) are
constant in time. It provides subsidiary constraints on
distribution functions $f^{(\alpha_i,\nu_i)}$. These conditions
assure that in a dynamical evolution the symmetry of the system is
preserved.

We now convert the global constraint into a time evolution
condition and consider:
\begin{equation} \label{cond}
\frac{d}{dt}\overline{P^{\Lambda,\lambda_{\Lambda}}_{\{N^{(\alpha_1)}_{\nu_{\alpha_1}},
\dots,\,N^{(\alpha_n)}_{\nu_{\alpha_n}}\}}}=0 .
\end{equation}
Introducing  here the result of Eq.\,(\ref{weights}) one obtains:
\begin{eqnarray}
&&0 = \frac{d\,{\mathcal A}^{\{{\mathcal N}\}}}{dt}d(\Lambda)\times\nonumber\\[4pt]
&&\int\limits_G d\mu(g)\bar\chi^{(\Lambda)}(g)
[D^{(\alpha_1)}_{\nu_1\nu_1}]^{N^{(\alpha_1)}_{\nu_{\alpha_1}}}\cdots
[D^{(\alpha_n)}_{\nu_n\nu_n}]^{N^{(\alpha_n)}_{\nu_{\alpha_n}}}\nonumber\\[4pt]
&&+ \sum_{j=1}^n\sum_{\nu_{\alpha_j}}\,
\frac{d\,N^{(\alpha_j)}_{\nu_{\alpha_j}}}{dt} {\mathcal
A}^{\{{\mathcal N}\}}d(\Lambda)\nonumber\\[4pt]
&&\times \int\limits_G d\mu(g)\bar\chi^{(\Lambda)}(g)
[D^{(\alpha_1)}_{\nu_1\nu_1}]^{N^{(\alpha_1)}_{\nu_{\alpha_1}}}\nonumber\\[4pt]
 &&\cdots
[D^{(\alpha_n)}_{\nu_n\nu_n}]^{N^{(\alpha_n)}_{\nu_{\alpha_n}}}
\log[D^{(\alpha_j)}_{\nu_j\nu_j}] .
 \label{deriv}
\end{eqnarray}
All integrals which appear in Eq.\,(\ref{weights}) and
Eq.\,(\ref{deriv}) can be expressed explicitly in an analytic form
for any compact symmetry group.

To write an expression for the time derivative of the
normalization factor ${\mathcal A}^{\{{\mathcal N}\}}$ we perform
analytic continuation from integer to continuous values of
variables $N^{(\alpha_n)}_{\nu_{\alpha_n}}.$ Thus we replace all
factorials by the $\Gamma$--function  of corresponding arguments.
We encounter here also the digamma function $\psi$ \cite{Abram}:
\begin{equation}\label{digamma}
\psi(x)=\frac{d\, \log\Gamma(x)}{d\,x} .
\end{equation}
This allows to write:
\begin{eqnarray}
&&\frac{d\,{\mathcal A}^{\{{\mathcal N}\}}}{dt}\nonumber\\[4pt]
&& = {\mathcal A}^{\{{\mathcal N}\}}\sum_j\sum_{\zeta_j}
 \left[\frac{d\,{\mathcal N}^{(\alpha_j)}_{(\zeta_j)}}{dt}
 \psi({\mathcal N}^{(\alpha_j)}_{(\zeta_j)}+1)\right.\nonumber\\[4pt]
 &&-\left.\sum_{\nu_{\alpha_j}}
\frac{d\,{\mathcal N}^{(\alpha_j)}_{\nu_{\alpha_j};(\zeta_j)}}{dt}
\psi({\mathcal N}^{(\alpha_j)}_{\nu_\alpha;(\zeta_j)}+1)\right] .
\label{evfactor}
\end{eqnarray}

The time derivatives ${d\,{\mathcal
N}^{(\alpha)}_{\nu_\alpha;(\zeta)}}/{dt}$ are obtained  from the
integrated kinetic equation fulfilled by a set of distribution
functions $f^{(\alpha_i,\nu_i)}_{(\zeta)}(\Gamma,\vec r,t)$. The
case of the generalized Vlasov - Boltzmann kinetic equations was
considered in \cite{TurRaf}.

These subsidiary conditions fulfilled by the microscopic kinetic
equations are the necessary conditions for preserving the internal
symmetry on the macroscopic level. Rates of change ${d\,{\mathcal
N}^{(\alpha)}_{\nu_\alpha;(\zeta)}}/{dt}$ are related to
``macrocurrents", which are counterparts of ``microcurrents"
related directly to a symmetry on a microscopic level via the
Noether theorem. This can be considered as a set of conditions on
macrocurrents to provide consistency with the overall symmetry of
the system. These conditions leads to nontrivial results only for
nonabelian symmetry groups. In the abelian case all charges are
additive ones. The charge conservation on the microscopic level is
equivalent to the global charge conservation of the multiparticle
system. This is not the case for nonabelian symmetries, where
nonabelian charges can combine to different representations of the
symmetry group.

New constraints on kinetic equations lead to decreasing number of
available states for the system during its time evolution. One can
expect that such a system when approaching its equilibrium would
produce less entropy. This should give some observable effects,
e.g. for particle production processes in heavy-ion collision.

It should be noted that these new constraints on evolution
equations are purely quantum effect. In the case of classical
systems a concept of representations of the symmetry group is not
applicable.

\pagebreak
\section*{Acknowledgments}
Work supported in part by the Polish Committee for Scientific
Research under contract KBN~-~2~P03B~030~18\,.

\end{document}